\documentstyle[preprint,prc,aps]{revtex}
\begin{document}
\preprint{RUB-MEP-81/94}
\draft
\title{ Meson-Exchange Currents in pp-Bremsstrahlung}
\author{J. A. Eden and M. F. Gari}
\address{ Institut f\"ur Theoretische Physik\\
	  Ruhr Universit\"at Bochum, D-44780 Bochum, Germany\\
	  {\rm email: jamie@deuteron.tp2.ruhr-uni-bochum.de}}
\date{22 July 1994; Revised 23 January 1995}
\maketitle
\tighten
\begin{abstract}
The long-standing discrepancy between pp-bremsstrahlung data and
calculations based on the relativistic impulse approximation current
is substantially reduced by the inclusion of the PV$\gamma$ and
intermediate-state
$\Delta$-resonance iso-scalar meson-exchange currents.
The success of the standard procedures adopted here shows that
pp-bremsstrahlung provides a powerful tool for the further study
of $T$=1 isoscalar exchange currents.
The need for addition high-precision data is stressed.
\end{abstract}
% %
\pacs{PACS:13.75.Cs, 25.20.Lj}
%\narrowtext
%
In earlier work \cite{JE93} we calculated a number of pp-bremsstrahlung
observables within a relativistic meson-baryon model. We included initial- and
final-state interactions, as well as the numerically cumbersome rescatter
contributions, as shown in Fig~1(a-c).  We placed special emphasis on a
consistent approach to both the 1- and 2-body problems, {\it i.e.}, the strong
interaction potential\cite{DP94,SD90} and the form factors it {
contains\cite{SD90,JF92}, were both microscopically calculated from a
consistently
parameterized meson field-theoretic Lagrangian. However, we considered the
electromagnetic interaction only in relativistic impulse approximation and our
results shared the discrepancy with experiment\cite{TE90} reported
elsewhere\cite{HF86,VB91,VH91,VH92,HF93a,HF93b,KA93}.

In Fig~2 we present results of these calculations, now extended to provide
the first fully consistent comparison of the Paris\cite{Paris},
Nijmegen\cite{Nijmegen}, Bonn\cite{Bonn} and RuhrPot\cite{DP94} interactions.
The selected geometries essentially correspond to best- and worse-case
comparison
with experiment.
It may be observed that, when the data is not dominated by uninteresting
phase space variations, the differences between the results from these models
is
smaller than their collective discrepancy with experiment. We conclude that the
failure of all recent attempts to describe the pp-bremsstrahlung data has
little
to do with the model-dependent choice of the half-off-shell $t$-matrices, but
indicates a sensitivity to the $T=1$ isoscalar meson-exchange currents.

Indeed, some 16 years ago Taitor {\it et al} \cite{LT78} showed that the
pp-bremsstrahlung observables at photon energies as low as 80~MeV may take
large contributions from meson-exchange currents involving an
intermediate-state
$\Delta$-resonance. Such results must be taken as indicative, rather than
conclusive, because the off-shell direct amplitudes shown in Fig~1(a-c) were
included as the measured elastic scattering cross section data.
This not only implies that the direct amplitudes are taken in the notoriously
unreliable soft-photon approximation, it also means that all interferences
between the meson-exchange and direct amplitudes are necessarily neglected.
Not even the relative sign of these contributions can be properly defined
and must simply be chosen to optimize the correspondence with experiment.
Nonetheless, the calculations by Taitor {\it et al} are useful and interesting
because they predicted a large contribution from the intermediate-state
$\Delta$-resonance exchange currents, and this motivates us to examine
such processes in a more systematic fashion.

In the present work we extend our earlier calculations to include the dominant
meson-exchange currents contributing to the pp-bremsstrahlung observables.
The exchange currents are calulated in the standard field-theoretic formalism
\cite{CH71,DR89,RW79} and include not only the intermediate-state
$\Delta$-resonance excitation via
$\pi-$ and $\rho-$meson exchange, as shown in Fig~1d, but also the
$\rho\pi\gamma$, $\omega\pi\gamma$, $\rho\eta\gamma$, $\omega\eta\gamma$
interactions, as shown in Fig~1e.
We adopt the RuhrPot\cite{DP94} interaction in order to obtain consistency
between the meson-exchange currents and the wave functions.
This immediately fixes all NN-meson coupling constants and form factors.
For the exchange currents involving intermediate-state $\Delta$-resonance
excitation we use,
\begin{eqnarray}
{\cal L}_{{\rm N}\Delta\pi} &=& -{g_{{\rm N}\Delta\pi}\over 2m}
 \bar{\psi}^\mu \vec{\tau}_{{\rm N}\Delta} \psi \partial_\mu \vec{\pi} + {\rm
h.c.}
\cr
{\cal L}_{{\rm N}\Delta\rho} &=& - i {g_{{\rm N}\Delta\rho}\over 2m}
 \bar{\psi}^\mu \gamma^5 \gamma^\nu \vec{\tau}_{{\rm N}\Delta} \psi
\vec{\rho}_{\mu\nu} + {\rm h.c.}
\cr
{\cal L}_{{\rm N}\Delta\gamma} &=& - i {e_p\mu_{{\rm N}\Delta}\over 2m}
 \bar{\psi}^\mu \gamma^5 \gamma^\nu \vec{\tau}_{{\rm N}\Delta}^{\,3} \psi
F_{\mu\nu} + {\rm h.c.}
\end{eqnarray}
and take $g_{{\rm N}\Delta\pi}$=28.85 to be consistent with the observed width
$\Gamma_{\Delta}$=115~MeV
(This value is also consistent with both Chew-Low theory\cite{BR75} and the
Strong Coupling Model\cite{GR76}).
We also adopt the familiar SU(6) result
$g_{{\rm N}\Delta\rho}$=$g_{{\rm NN}\rho}(1+\kappa_\rho)g_{{\rm
N}\Delta\pi}/g_{{\rm NN}\pi}$,
so that vector dominance implies
$\mu_{{\rm N}\Delta}$ = ${1\over 2}(1+\kappa^{{\rm iv}}) g_{{\rm
N}\Delta\pi}/g_{{\rm NN}\pi}$.
For the real photon, the Dirac and Pauli electromagnetic form factors
obviously reduce to their normalization values, and since the present work is
confined to
low $Q^2$, we approximate the strong form factors as
$F_{{\rm N}\Delta\alpha}$ = $F_{{\rm NN}\alpha}$ for $\alpha$=$\pi,\rho$.
For the electromagnetic coupling to the vector-pseudovector meson currents we
have,
\begin{eqnarray}
{\cal L}_{{\rm PV}\gamma} &=& -{e_p g_{_{\rm PV}\gamma}\over 2m_{_{\rm V}} }
 \epsilon^{\mu\nu\sigma\tau} F_{\mu\nu} \vec{\phi}_{\sigma}^{\,^{\rm
V}}.\partial_\tau \vec{\phi}^{\,^{\rm P}},
\end{eqnarray}
with V=$\rho,\omega$ and P=$\pi,\eta$, and we fix the coupling constants to
their
experimental \cite{PDG} values of
$g_{\rho\pi\gamma}$=0.55,
$g_{\omega\pi\gamma}$=2.03,
$g_{\rho\eta\gamma}$=1.39 and
$g_{\omega\eta\gamma}$=0.31.
We emphasize that this description of the meson-exchange currents introduces no
free parameters and is fully consistent with both the strong-interaction and
the form factors it contains. Since we retain our microscopic description of
the
NN-interaction, all interferences are included. No form of soft photon
approximation
is adopted at any stage.

There are important differences between the present work and other recent
calculations which incorporate N$\Delta\gamma(\pi,\rho)$, $\rho\pi\gamma$,
$\omega\pi\gamma$, $\rho\eta\gamma$ and  $\omega\eta\gamma$ currents into
pp-bremsstrahlung.
First, like ref~\cite{MJ94}, we retain the dominant $VP\gamma$ exchange
currents that
are neglected in ref~\cite{JO94}. These currents are certainly smaller
than the N$\Delta\gamma(\pi)$ exchange currents but, in particular,
the $\omega\pi\gamma$ contribution has a magnitude which is similar to that
of the N$\Delta\gamma(\rho)$.
Second, we require no unphysical free parameters to avoid double-counting
of N$\Delta$ intermediate states and our approach guarantees a consistent
description of coupling constants, form factors and particle masses in both
the exchange currents and the wave functions.
Finally, although the $t_{\Delta{\rm N}}$-matrix expansion carries the
advantage of
providing a non-perturbative description of the $\Delta$ degrees of freedom in
the wave function, it also creates the necessity to compute wave function
re-orthonormalization contributions. We realize that the well known
cancellation of
wave function re-orthonormalization and meson-recoil amplitudes in the
NN$\rightarrow$NN
transition \cite{GA76} fails for NN$\rightarrow\Delta$N transitions, even in
the
static limit. As such, the neglect of both meson-recoil and wave function
re-orthonormalization contributions in refs~\cite{MJ94,JO94} intoduces an
approximation whose importance is difficulty to assess in the presence of a
free parameter.

By describing the exchange currents perturbatively, we automatically include
the meson-recoil contributions and we formally have no need of
wave function re-orthonormalization terms \cite{GA76}. In addition, the
precision of this
perturbative approach can be crudely estimated by noting that the
impulse approximation rescatter amplitudes of Fig~1c represent a correction
of less than 15\% to the dominant nucleon-pole contributions shown in
Fig.~1(a,b).
At the geometries where comparison is possible, the numerical results of our
exchange current calculations are actually fairly similar to the $t_{\Delta{\rm
N}}$-matrix
calculations reported in refs~\cite{MJ94,JO94}, although we realize this
conclusion depends on the free parameter adopted in those works.

In Fig~3 we compare the individual contributions of the impulse approximation
current,
the perturbative intermediate-state $\Delta$-resonance excitation via $\pi-$
and
$\rho-$exchange, and $\rho\pi\gamma$, $\omega\pi\gamma$, $\rho\eta\gamma$ and
$\omega\eta\gamma$ exchange currents.
We fix the coplanar kinematics for this analysis at E$_{\rm lab}$=280~MeV with
$\theta_1=16.0^\circ$ and $\theta_2=27.8^\circ$ since the existing data at
this geometry exhibits significant variation from phase space and is typical
of other data that are poorly described by recent calculations.
We observe that the $\omega\pi\gamma$ contribution is clearly larger than those
of the $\rho\pi\gamma$ , $\rho\eta\gamma$ and $\omega\eta\gamma$, the last two
being
essentially negligible.
Clearly the most important exchange current contribution involves
intermediate-state
$\Delta$-resonance excitation via $\pi$-exchange.

In Fig~4 we compare pp-bremsstrahlung observables with calculations that
include or neglect intermediate-state $\Delta$-resonance excitation and
$\rho\pi\gamma$, $\omega\pi\gamma$, $\rho\eta\gamma$ and $\omega\eta\gamma$
interactions.  Clearly the inclusion of the exchange currents substantially
reduces the long-standing discrepancy with experiment.
We checked all other geometries reported in ref~\cite{TE90} and found
essentially the same result.
(The data are presented here without the arbitrary ${2\over 3}$ scaling that
has sometimes been introduced to improve the correspondence to the impulse
approximation results).
The magnitude of the exchange current contributions shows, contrary to the
expectations of earlier publications, that pp-bremsstrahlung cannot be simply
calculated in impulse approximation to test the accuracy of various off-shell
NN $t$-matrices. The obvious consequence of this is that the process can only
be
meaningfully investigated within a microscopic theory where consistency between
the
strong form factors, the NN-interaction and the exchange currents can be
explicitly
guaranteed.

We stress that both the {\it sensitivity} and {\it selectivity} of the
pp-bremsstrahlung
data to the $T=1$ isoscalar meson-exchange currents provides a unique
opportunity for the
study of a range of more subtle effects of current interest. These include not
only
resonance width effects, but also relativistic corrections to the exchange
current operators.
To that end, we appeal to the experimental community to provide more high
precision data, particularly at the kinematical conditions where the impulse
approximation
calculations fail.

In conclusion, we find that a consistent and microscopic inclusion of
the dominant meson-exchange current contributions in pp-bremsstrahlung
shows that the application of standard calculation procedures substantially
reduces the long-standing discrepancy with experiment.
No free parameters enter the present calculation. Consistency between the
wave functions and exchange currents is guaranteed since the
the NN-interaction, the strong form factors and all meson-exchange currents
are all calculated from a consistently parameterized effective Lagrangian.
All interferences have been calculated without approximation and no form
of soft photon approximation is used at any stage. Additional high-precision
data will be critical to the further study of resonance width and relativistic
effects,
provided emphasis is placed on the kinematically complete geometries where the
impulse
approximation results are at variance with experiment.

\acknowledgements
This work is supported by COSY-KFA J\"ulich (41140512).

% REFERENCES

% FIGURE CAPTIONS
%Fig 1
\begin{figure}
\caption{Dominant contributions to pp-bremsstrahlung observables:
(a) initial-state, (b) final-state and (c) rescatter impulse contributions,
(d) intermediate-state $\Delta$-resonance and (e) $\rho\pi\gamma$,
$\omega\pi\gamma$, $\rho\eta\gamma$ and $\omega\eta\gamma$ meson-exchange
currents.
All such contributions are included in the present work.}
\end{figure}
%
%Fig 2
\begin{figure}
\caption{
Comparison of coplanar pp-bremsstrahlung data [5] at E$_{\rm lab}$=280~MeV
with calculations using wave functions obtained from various NN-interaction
models but current densities confined to the relativistic
impulse-approximation.
Initial-, final- and rescatter-correlations are included with partial waves
summed to $J_{\rm max}$=8.
All models give essentially equivalent results which differ from experiment
when phase space variations are not extreme.
}
\end{figure}
%
%Fig 3
\begin{figure}
\caption{
Square of the pp-bremsstrahlung invariant amplitude as a function of
the photon emission angle at $E_{\rm lab}=280$~MeV
for proton co-planar scattering angles of 16.0$^\circ$ and 27.8$^\circ$.
The impulse approximation (IA) corresponds to Fig~1(a-c).
The $\pi-$ and $\rho-$exchange contributions to the N$\Delta\gamma$ exchange
currents of Fig~1(d) are shown separately. The $\rho\pi\gamma$ and
$\omega\pi\gamma$ correspond to Fig~1(e).
The negligible amplitudes resulting from the $\rho\eta\gamma$ and
$\omega\eta\gamma$ contributions are not shown.
}
\end{figure}
%
%Fig 4
\begin{figure}
\caption{
Comparison of coplanar pp-bremsstrahlung data [5] at E$_{\rm lab}$=280~MeV
with calculations using wave functions obtained from the RuhrPot NN-interaction
and current densities that include (IA+MEXC) or neglect (IA) the
N$\Delta\gamma$,
$\rho\pi\gamma$, $\omega\pi\gamma$, $\rho\eta\gamma$ and $\omega\eta\gamma$
exchange currents. The long-standing discrepancy with experiment is
substantially
reduced at these and other geometries.
}
\end{figure}

% TABLES
%\begin{table}
%\end{table}
\end{document}